\documentclass[prb,twocolumn,amsmath]{revtex4}
\usepackage{graphicx}

\begin{document}

\title{Heavy anion solvation of polarity fluctuations in Pnictides}

\author{G.A. Sawatzky}
\affiliation{Department of Physics and Astronomy, University of British Columbia, Vancouver, BC V6T-1Z1, Canada}

\author{I.S. Elfimov}
\affiliation{Advanced Materials and Process Engineering Laboratory, University of British Columbia, Vancouver, BC V6T 1Z4, Canada}

\author{J. van den Brink}
\affiliation{Institute Lorentz for Theoretical Physics, Leiden University, P. O. Box 9506, 2300 RA Leiden, The Netherlands}
\affiliation{Institute for Molecules and Materials, Radboud Universiteit Nijmegen, P.O. Box 9010, 6500 GL Nijmegen, The Netherlands}

\author{J. Zaanen}
\affiliation{Institute Lorentz for Theoretical Physics, Leiden University, P. O. Box 9506, 2300 RA Leiden, The Netherlands}


\maketitle

{\bf Once again the condensed matter world has been surprised by the discovery of yet another class of high temperature superconductors \cite{kamihara-jacs-08,wen-epl-08,ren-epl-08,cheng-08,chen-prl-08,takahashi-nature-08,chen-nature-08,rotter-arXiv-08,chen-arXiv-08}. The discovery of iron-pnictide  (FeAs) and chalcogenide (FeSe) based superconductors with a $T_c$ of up to 55 K is again evidence of how complex the many body problem really is, or in another view how resourceful nature is. The first reactions would of course be that these new materials must in some way be related to the copper-oxide based superconductors for which a large number of theories exist although a general consensus regarding the correct theory has not yet been reached. Here we point out  that the basic physical paradigm of the new iron based superconductors is entirely different from the cuprates. Their fundamental properties, structural and electronic, are dominated by the exceptionally large pnictide polarizabilities.} 

{\it Cuprate essentials.}
There is little doubt that the electronic properties of copper-oxide parent compounds are determined by the strong Cu $3d$ on site Coulomb repulsions and moderate O $2p$ to Cu $3d$ charge transfer energies, leading to energy gaps for charge fluctuations of 2 eV or greater, localized spin 1/2 magnetic moments on Cu, extremely strong superexchange driven antiferromagnetic coupling of these localized spins and magnetic excitation energy scales for single magnons extending up to 0.5 eV. 

\begin{figure*}
\centering
\includegraphics[clip=true,width=.8\textwidth]{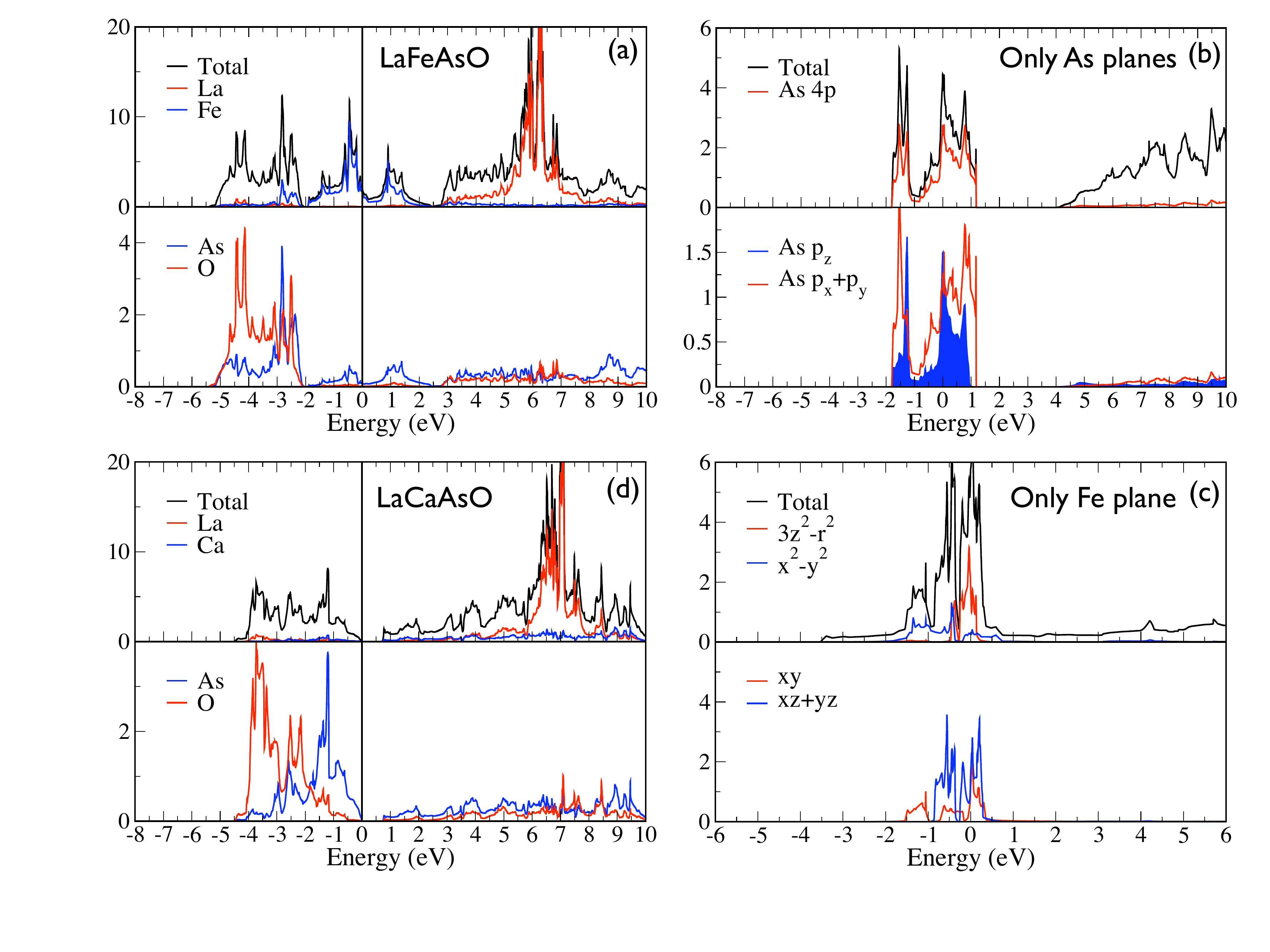}
\caption {LDA density of states for (a) LaFeAsO; (b) As and (c) Fe planes calculated 
leaving out all the other atoms; (d) LaCaAsO.  The zero of energy is at the Fermi energy, $E_f$.}
\label{all-DOS}
\end{figure*}

Another often quoted important property of the cuprates is the two dimensional character of the magnetic interactions and the electrical transport properties. This 2D behavior strongly enhances the importance of magnetic quantum fluctuations as evidenced by the strong zero point spin deviations and suppression of the long range magnetic ordering temperature ($T_N$) from that expected from mean field theory. Another undisputed fact is the strong degree of covalent mixing of the O 2$p$ orbitals and the Cu 3$d$ states in the cuprates. This drives the very strong superexchange interactions, and the reduction of the local moments. This quite naturally leads to the conclusion the electron lattice interactions can be very strong leading according to some to the importance of electron-phonon coupling in describing the transport properties of these materials.
 
{\it Comparing cuprates and pnictides: local picture.} 
So what are really the similarities and differences between the cuprate high $T_c$'s and the iron-pnictide FeAs and iron-chalcogenide FeSe based compounds? First of course both Cu$^{2+}$ and Fe$^{2+}$ (in its high spin state) have strong tendencies to form local magnetic moments. In both cases band theory predicts a rather two dimensional behavior at low energy scales\cite{pickett-rmp-89,pickett-prb-90,andersen-prb-94,lebegue-prb-07,singh-prl-08,xu-epl-08,yin-prl-08}. For Cu$^{2+}$ in a square planar coordination of oxygens the local spin is 1/2.  For Fe$^{2+}$ in a tetrahedral coordination of As (or Se), which inverts the energies of the $e_g$ and $t_{2g}$ orbitals, one would expect a local spin of 2 and an orbital degeneracy because of the singly occupied minority spin $e_g$ state. This is all with the assumption that in both systems the on site $d$-$d$ Hubbard U is much larger than the $d$ band width and in the case of Fe the Hund's rule coupling is larger than crystal and ligand field splitting of the Fe $d$ levels resulting in a high spin case. We have also assumed that the charge transfer energy in the ZSA scheme\cite{ZSA} is positive in both cases resulting in a charge excitation gap. 

At first glance though the case for Fe being in this local moment high spin state is much less convincing than it is for Cu. First of all we are dealing with arsenides and generally we would expect the effects of covalency on the ligand field splitting to be so large that Hund's rule would be overruled and a low spin state with spin 1 would result. A further distortion from T$_d$ point group symmetry could split the $t_{2g}$ states which in principle could result in a spin zero state --which would however require quite a large distortion. 

For the cuprates it is generally accepted that the gap is of a charge transfer type because the Cu $d$-$d$ interaction (Hubbard $U$) is much larger than the charge transfer energy and the charge transfer energy for transition from O $2p$ to a Cu $3d^{10}$ state is positive. This is a much less clear scenario for the arsenides. First the electron negativity of As$^{3-}$ is much smaller than for O$^{2-}$ so in this ionic picture we would again expect to be in the 
charge transfer regime of ZSA\cite{ZSA} but the charge transfer gap should be very small and probably negative, resulting in a $p$-type metal. 

The magnetic properties of LaFeAsO, although somewhat anomalous, are not at all suggestive of a local moment S=2 state but rather suggest at most a small local moment and a temperature independent Pauli susceptibility enhanced somewhat from a band structure prediction\cite{kamihara-jacs-08,singh-prl-08} and a transition to an antiferromagnetic spin density wave phase below about 150K\cite{cruz-nature-08}. Of course, a low spin state can only be achieved if the Hund's rule coupling $J_H$=(1/14)(F$^2$+F$^4$) where F$^2$ and F$^4$ are Slater type integrals, is strongly reduced to a fraction of its atomic value and becomes comparable to the small crystal field splitting predicted by band theory\cite{haule-prl-08,boeri-prl-08,ma-arXiv-08}. This however is hard to realize because as discussed in the past\cite{antonides-1-prb-77} the higher order Slater integrals are not reduced by a polarizable surrounding to lowest order because they do not involve a charge but rather a quadrupole or dipole moment so the reaction of the surroundings will cause small changes in these. On the other hand the monopole integral F$^0$ which basically determines the Hubbard $U$ will be strongly influenced by the surroundings because it dictates the energies involved in actual charge fluctuations involving only $3d$ electrons. 

{\it Iron-pnictide bandstructure effects.}
Up to here we have limited our discussion to a local approach neglecting the translational symmetry and band structure effects. These local approaches are very successful in discussing the basic aspects of the electronic structure of many 3$d$ transition metal oxides. However, band structure effects and translational symmetry is another important part of the problem. In a translational invariant solid the tendency  to form local moments is first of all dictated not by Hund's rule coupling but rather by the competition between the Hubbard $U$ which suppresses on-site polarity fluctuations and the one-particle band dispersion which lives off polarity fluctuations. So the first condition for the formation of a local moment is that $U>W$ where $W$ is the one electron $3d$ band width. Recent LDA calculations however indicate a $d$ band dispersion width of at most about 2eV so we need to find a way of reducing the Hubbard $U$ to about 2eV in order to suppress local moment formation. 

{\it Arsenic bandwidth and its spead-out charge distribution.}
This brings us to the heart of the discussion of the large differences between oxides and arsenides. There are several things that are of central importance here. First of all in the oxides the O 2$p$ states form rather large band widths of about 6eV in the cubic oxides and lowered somewhat to 4eV in the two dimensional cuprates. The LaFeAsO layer structure is however very different from the closed packed O lattices in most oxides. In fact the As-As bond length of almost 4 \AA~is extremely large. So in spite of the large orbital radii of the 4$p$ electron states the As-As hopping integrals are quite small. This is nicely demonstrated by a bandstructure calculation of the As lattice in the LaFeAsO structure, while leaving out all the other atoms, see Fig.\ref{all-DOS}(b). The total arsenic bandwidth is only about 3eV so that As in this system behaves like a narrower bandwidth system than O does in oxides. 

Having concluded on the rather atomic like nature of the role played by As, we also observe that the As charge density is in fact strongly spatially spread out. We will come back to this later. We first look at the role of the Fe-As hybridization. The total and atom projected density of states are shown in Fig.~\ref{all-DOS}(a). As also others have shown\cite{lebegue-prb-07,singh-prl-08,xu-epl-08,yin-prl-08,ma-arXiv-08} the only bands prominently present close to the chemical potential are Fe 3$d$ based bands. In fact looking at the partial projected density of states the amount of As character in this energy region is rather small. 
Of course this observation is qualitative since the partial density of states refers to states inside the muffin tin region and since as mentioned above the As 4$p$ states are rather extended we may not be capturing their full contribution. Nonetheless comparing the Fe 3$d$ density of states close to $E_f$ with the total, which involves all the states also outside the muffin tin radii we see that these states are very strongly Fe 3$d$ dominated. To demonstrated this rather modest Fe-As hybridization in this energy region we show in Fig.~\ref{all-DOS}(c) the Fe density of states we get if we remove all the other atoms from the structure. This is the band structure that would result from only Fe-Fe hoping integrals. We see that this density of states and total band width is comparable to that of the compound itself. 

\begin{figure}
\centering
\includegraphics[clip=true,width=0.4\textwidth]{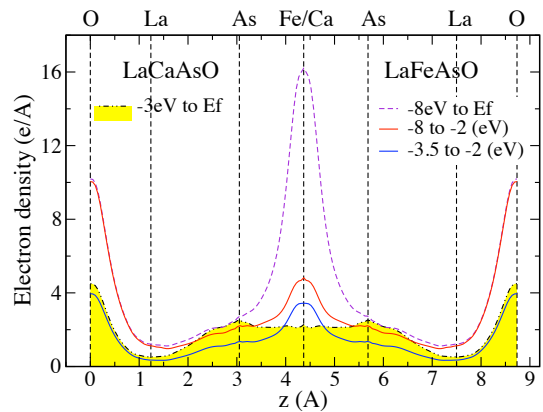}
\caption {Electron charge distribution in the unit cell of LaFeAsO (a) and LaCaAsO (b) as a 
function of z calculated with various energy integration limits and integrated over the ab-plane. 
The atomic planes are depicted with vertical dashed lines.}
\label{charge-z-Fe}
\end{figure}

All this leads us to a rather surprising picture of basically an Fe $3d$ band structure with little hybridization with As and a dispersional width of about 2eV plus modest crystal and ligand field effects resulting in additional $3d$ band splittings. But at the same time the electrons in this Fe $d$ band are propagating in the close proximity of a rather extended charge distributions around the As centers. It is this extended charge distribution about the As in which the Fe 3$d$ electrons have to move that interests us most at this time. To demonstrate the spatial extend of the As charge distribution further we show the occupied charge density integrated over the $ab$ plane and integrated over energy between the limits as indicated in the Fig.~\ref{charge-z-Fe}(a) as a function of $z$, the distance along the $c$ direction. Indicated also in the figure are the positions of the As, Fe, La, and O planes. Integrating over the energy from -8eV to $E_f$ we see a strong peak at the Fe plane due to the Fe 3$d$ electron states sitting on a rather broad distribution of electron density extending far past the As plane but dropping to almost zero at the La plane position. This broad distribution is a result of the As based electrons which have a very broad rather constant density extending far away from the As nuclear position towards the Fe and La planes. The charge density in the region -3.5 to -2 eV is again this broad distribution in the As plane region. This demonstrates again the highly peaked density caused by the Fe $d$ states positioned in and close to a rather broad density resulting mostly from As $4s$, $4p$ and $5s$-like states. 

To demonstrate this even further we looked at the fictitious compound LaCaAsO in which the $3d$ states below the Fermi energy are absent. The same lattice parameters as for the Fe compound were used and the density of states is shown in Fig.~\ref{all-DOS}(c). The region close to $E_f$ is again dominated by As states now hybridized with the Ca $4s$ and $4p$ states which are mainly above $E_f$. The system has a band gap of 0.7eV in LDA. Important for our discussion is the charge density calculated in the same way as above as a function of the distance along the $c$ axis which for the energy range -3 to $E_f$ clearly demonstrates this broad charge distribution resulting from the As valence states hybridized with the Ca $4s$ and $4p$ bands. 

So the picture that emerges is that of $d$ electrons moving in a broad distribution of charge density which is rather loosely bound to but resulting from As 5$s$, 4$p$ states hybridized with Fe 4$s$, 4$p$ states. This is the kind of charge distribution one would get from a sparse lattice of large negative As  ions with spatially extended $4p$ orbitals strongly penetrating into the interstitial region between Fe and As and also between La and As. The large As-As bond length results in a small As $4p$ band dispersion. The large spatial extend of these $4p$ orbitals is to a large extend caused by the fact that a $4p$ orbital has two radial nodes. Also the tail region is much less strongly angularly directed than the O $2p$ orbitals. So inspite of the large radial extend the hybridization with the strongly directed Fe $d3$ orbitals is rather limited. 

The model that presents itself is the one of tightly bound Fe $3d$ electrons propagating in a lattice of rather isolated As negative ions. This however raises immediately a number of questions such as: (i) What stabilizes this strange structure in the first place? (ii) Why is there not a large local Fe moment in spite of the rather narrow $d$ bands and the rather atomic nature of the 3$d$ states? (iii) What could possibly be the origin of effective attractive interactions between the $d$ electrons resulting in a superconducting state? 

{\it What stabilizes the layered pnictide structure?} 
Conventional wisdom tells that the Madelung potential plays an important role in the cohesive energy of ionic compounds like the transition metal oxides. The Madelung potential has its largest contribution in rather cubic structures in which the positive cations are surrounded by negative anions and vise versa. Layered structures on the other hand break this rule because they make use of the large polarizability of one of the ions, either the anion or cation. The most common of course are the sulfides, selenides and tellurides such as TiS$_2$, MoS$_2$, TiTe$_2$ etc. but there are also numerous examples in halides like the bromides and iodides. As discussed in detail by Haas\cite{haas-81} and by Wilson and coworkers \cite{wilson-jpcm-94} the layer structures make use of the strong polarizability of usually the anion which in the structure is placed in a very asymmetric surrounding of cations which then results in a large electric field at the anion site. This polarizes the anion and the induced dipoles interact with neighboring  dipoles and the monopoles of the positive central layer to lower the energy. If the polarizability is large enough then this induced dipole-monopole interaction can be larger than the point charge interactions in a cubic lattice. 

In the most common layer structure a Ti$^{4+}$ central layer is sandwiched between two S$^{2-}$ layers providing the conditions just described. These sandwiches are then repeated. Another very important aspect of these kinds of layer structures is that the surface consisting of charged S$^{2-}$ ions are in spite of the charge {\it not} polar since the charge is exactly -1/2 of the charge of the central layer underneath which is the condition for a non-polar surface. Polar surfaces will reduce their large energy due to internal electric fields by reconstructing and one way is to electronically reconstruct by moving electrons so that the surface layer has -1/2 the charge of the layer underneath\cite{hesper-00,ohtomo-04}.  In the definition of a truly layered compounds we should  include the condition of a non-polar surface when cleaving parallel to the layers. With this definition materials like YBCO for example are not really layered since such surfaces would always be polar\cite{hossain-nphys-08}. Also LaFeAsO is not really layered in this sense and we would guess would not be easily cleavable. The reason is that the Fe$^{2+}$ is surrounded by 1/2 layers of As$^{3-}$ yielding a net charge of $-1$ for this sandwich which in the $c$ axis direction then alternates with a charge of $+1$ for the LaO sandwich. The way to stabilize this structure though is to make use of the extremely large polarizability of As as well as the substantial polarizability of La. 

Recall that the polarizability $\alpha$ of ions is roughly equal to their volume. So the polarizability of O$^{2-}$ is between 0.5 and 3.2 \AA$^3$ ~\cite{tessman-pr-53,dimitrov-jap-96,shannon-prb-06} and obviously that of S$^{2-}$ (4.8 - 5.9 \AA$^3$), Se$^{2-}$ (6 - 7.5 \AA$^3$) and Te$^{2-}$ (8.3 - 10.2 \AA$^3$) are progressively larger. This is why oxides mostly form cubic structures and the heavier chalcogenides form layered structures\cite{haas-81}. Similarly, fluorides form cubic structure and chlorides, bromides and iodides often layered ones. The polarizability of As$^{3-}$ can easily be derived from the ones of the chalcogenides. The radius of As$^{3-}$ (222 pm) is  comparable but large than that of Te$^{2-}$ (211 pm) and Se$^{2-}$ (191 pm), and according to the volume ratios As$^{3-}$ has a polarizability of  9 - 12 \AA.
Actually, also La$^{3+}$ has a reasonably large polarizability of 3 to 6 \AA$^3$ i.e. larger than O$^{2-}$ \cite{shannon-prb-06,shannon-jap-93}. These large polarizability stabilize the  As-Fe-As sandwich structure, to which the La-O-La sandwich is slaved.

\begin{figure}
\centering
\includegraphics[clip=true,width=0.4\textwidth]{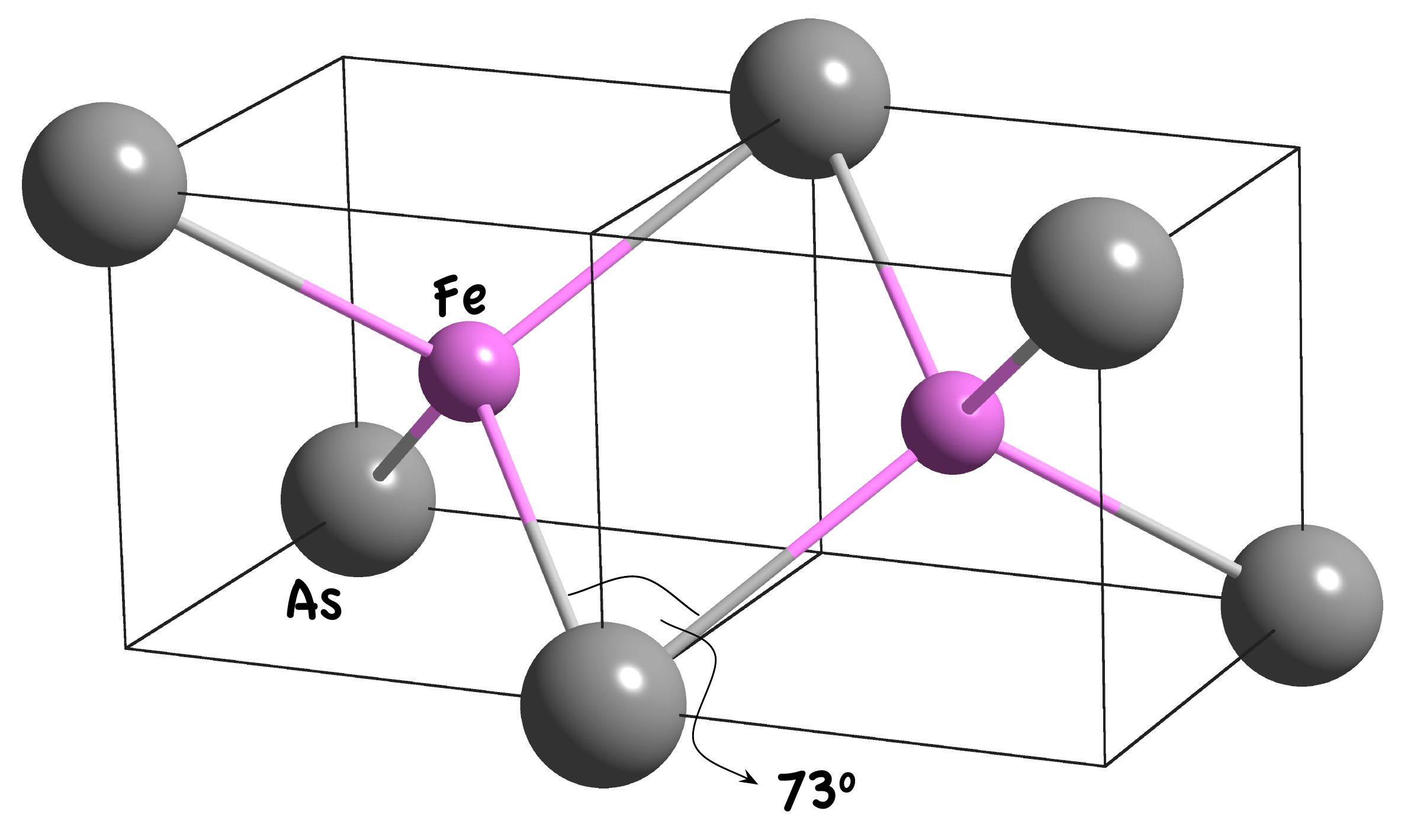}
\caption {FeAs tetrahedral sub-lattice in the crystal structure of LaFeAsO formed by the edge sharing FeAs$_4$ tetrahedra.}
\label{FeAs-str}
\end{figure}

{\it Screening of Hubbard U.}
What about the small or non existent local moment in spite of the relative narrow Fe 3$d$ band widths? Again here the polarizability of the As can play a prominent role. As described in our studies of the oxides in the mid 1980's\cite{deBoer-prb-84}, the reduction of the Hubbard $U$ from its atomic value of about 20 eV to less than 10 eV in oxides is to a large extend due to the polarizability of O$^{2-}$. The Hubbard $U$ is formally defined as the additional energy it costs to place two electrons (or holes) on the same site relative to having them on far removed sites, which is equivalent to the ionization potential minus the electron affinity of an atom with a given number of $d$ electrons (6 in the present case). Note that this involves a change of two particles. Now the ionization potential in a polarizable medium is lowered not by conventional screening effects described by a dielectric constant but by the polarization of the surroundings and the subsequent potential produced by these induced dipoles at the position of the ionized site. This is the same effect that stabilizes ions in a polar solvent --the solvation effect.

Neglecting first the dipole dipole interactions and assuming linear response theory we have $U=U_0-2E_p$ with polarization energy $E_p \sim \frac{1}{2}\sum_i \alpha_i E_i^2$ where $U_0$ is the bare free ion value and $E_i$ is the electric field at the $i$th site due to a charge of $\pm 1$ at the origin. If we neglect the dipole-dipole interactions and take the field $E_i = \pm{e}/{R_i^2}$ from the point charge and considering only the nearest neighbors we get $U=U_0-2(\frac{Z\alpha}{2}{e^2}/{R^4})$ where $Z$ is equal to the number of nearest polarizable neighbors. Using $\alpha=10$\AA$^3$ for the As$^{3-}$ polarizability, the Fe-As distance $R=2.4$\AA~ and $Z=4$ we obtain $U=U_0-17.3$eV! The very large induced dipole moments cause this enormous polarization energy. The size of these moments implies at the same time that dipole-dipole interactions cannot be neglected, and we find that in this case they reduce the $E_p$ of the four neighbors by about a factor three. As the Coulomb interaction is long range we also take the effects of further neighbors into account, which in the end leads to a total polarization of $2E_p=10.0$ eV, calculated for a plane of $4 \cdot 10^4$ polarizable As ions. In other words $U$ is strongly "solvated'', especially if in addition also the Fe-As bond polarizability is taken into account. So it is not so surprising that indeed $U$ is smaller than $W$, leading to a quenching of the local moment due to polarity fluctuations in the band theory. 

This large polarization reduction of the single particle ionization energy will result in the dressing of a particle much like that with phonons but now with virtual electron-hole excitations describing the polarizability of As. 
This would increase the effective mass of the particles although moving the particle to a nearest neighbor Fe site would not require the complete destruction of the previously build up polarization, as in a Holstein model of electron-phonon coupling. This problem is being looked into with a theoretical model and will be published in a separate paper\cite{mona-08}.  This effective mass enhancement will increase the Pauli like susceptibility from its LDA value. 

{\it Longer-range effects of large pnictide polarizability.}
What about an attractive interaction that could be responsible for the superconductivity? The on-site Coulomb interaction will most likely not be reduced to zero or become attractive. However there is something very special about this particular pnictide crystal structure that can strongly reduce the nearest neighbor interaction. The point is that two neighboring Fe ions have two common As ions as neighbors (Fig.~\ref{FeAs-str}). So let us go back to our original derivation of the polarization reduction of $U$ and do this for the nearest neighbor interaction $V$. One now obtains 
\begin{align*} 
   V = V_0 - \frac{1}{2} \sum_{common} \alpha [({\bf E}_1+{\bf E}_2)^2 -E_1^2-E_2^2] ,
\end{align*}
which reduces to  $V=V_0- 2 \alpha {\bf E}_1 \cdot {\bf E}_2$, where 2 refers to the number of common As neighbors and $E_1$ ($E_2$) are the electric 
fields of Fe$_1$ and Fe$_2$ with common As neighbors. This leads to $V = V_0 - {2\alpha e^2} (\cos \Theta)/ {R^4}$ where $\Theta$ is the Fe-As-Fe bond angle. For $\Theta=180^\circ$ the screening induced interaction is repulsive, as we showed in previous papers dealing with polarizability explicitly\cite{brink95,meinders95}. However for an angle of less than 90$^\circ$ this interaction is attractive and reduces the nearest neighbor Coulomb repulsion. Taking the polarization of the As ions in the plane into account and including the effect of dipole-dipole interactions, we obtain $V=V_0-3.83$eV. Now $V$ could easily become attractive since the bare repulsion between the Fe sites at a separation of 2.8 \AA~is only 5.1 eV and $10$ \AA$^3$ is a lower end estimate for the polarizability of As$^{3-}$. In the computation of the polarization energy above we find that due to the local field effects for $\alpha = 12$ \AA$^3$ the nearest neighbor Coulomb interaction becomes over-screened, so that effectively $V$ becomes attractive. 

These estimates demonstrate that it is quite possible because of the curious structure of the rather isolated As ions with a huge polarizability and the less than 90$^\circ$ Fe-As-Fe bond angle that the nearest neighbor $d$-$d$ Coulomb interaction becomes attractive, while longer range interactions are still repulsive. This could lead to $s$,$p$,or $d$ wave superconductivity, the lowest energy of which would be dictated by details of the band structure close to $E_f$ and also by the fact that a central Fe ion and two of its nearest neighbors share one As neighbor. This will reduce the effective mass for a motion in the placket around the central Fe ion and influence the energy of such a $s$,$p$,$d$ wave pair\cite{mona-08}. 

This mechanism of superconductivity reminds us clearly of the Little model put forward for organics\cite{little-pr-64}. In this model the phonons are basically replaced by virtual excitons on side chains to the chain carrying the free carriers. The role of the excitons is then the same as that of the phonons except now with a very large " phonon" energy which enters as a pre-factor in the BCS theory of Tc. In our case the polarizability of As can clearly be modeled with an electric field mixing the As $p$ and $s$ orbitals which one can think of as a virtual excitation of a $p$-$s$ exciton in the As "side" slab now rather than a chain. In previous work we have, in fact, modeled the polarizability of O$^{2-}$ in terms of $p$-$s$ virtual excitations on oxygen\cite{deBoer-prb-84}. This mechanism also reminds us of the one introduced by Allender, Bray and Bardeen at the interface between a semiconductor and a metal\cite{allender-prb-73}. Also here the virtual excitations in the semiconductor across the band gap or excitons could mediate an effective attractive interaction between the metal electrons. 
In our case the semiconductor is formed by As $4p$ bands as the valence bands and As $5s$ hybridized with Fe $4s$, $4p$ as conduction bands. The metal is formed by the Fe $3d$ bands.
Ginzburg was perhaps the first actually to discuss this type of mechanism\cite{ginzburg}. 

{\it Conclusions.} A model for superconductivity in the arsenides emerges that is a result of the strong dynamic polarizability of the large As$^{3-}$ ions. It leads to the Fe 3$d$ electrons propagating like electronic polarons, attaining a nearest neighbor attractive interaction. The peculiar pnictide lattice structure makes this interaction attractive because of the less than 90$^{\rm o}$ Fe-As-Fe bond angle. 
Notabene, it was recently discovered that FeSe forms a pnictide structure while it 
superconducts\cite{hsu-arXiv-08}, vividly illustrating the case that 
if we can reduce the chalcogen band width in going from the common
dichalcogen structure to the pnictide structure isolating the polarizable
chalcogen anion a superconducting state can arise.
This also suggests that this mechanism could very well be found in more materials with heavy, highly polarizable anions provided that they are dilute enough not to form very wide bands and that the bond angles meet the criteria described above. This actually leaves open a very wide range of possible new high temperature superconductors. 

We gratefully acknowledge 
Kristjan Haule, Andrea Damascelli and Doug Bonn for enlightening discussions; O.K. Andersen for providing unpublished downfolding band structure results; Mona Berciu for providing us with the first results of the polarizability model calculations and the electronic polaron.
This research was funded in part by the Canadian funding agencies NSERC, CRC, and CIFAR
as well as the Netherlands Foundation for Physical Research (FOM).

\end{document}